\documentclass[12pt]{iopart}  
\usepackage{iopams}
\expandafter\let\csname equation*\endcsname\relax
\expandafter\let\csname endequation*\endcsname\relax
\usepackage{amssymb}
\usepackage{amsmath}
\usepackage{mathrsfs}
\usepackage{graphicx}
\usepackage{flushend}
\usepackage{subfigure}
\usepackage{xcolor}
\usepackage[normalem]{ulem}
\usepackage{lipsum}
\usepackage[colorlinks, linkcolor=blue, anchorcolor=blue, citecolor=blue, urlcolor=blue]{hyperref}
\usepackage{tikz}
\usepackage{booktabs}
\usepackage{lineno}

\begin{document}
\title{The motion of twisted particles in a stellar gravitational field}
\author{Dan-Dan Lian, Peng-Ming Zhang\footnote{Corresponding author.}}
\address{School of Physics and Astronomy, Sun Yat-sen University, 519082 Zhuhai, China}
\ead{zhangpm5@mail.sysu.edu.cn}
\begin{abstract}

In this work, we explore the motion of a twisted particle possessing intrinsic orbital angular momentum (OAM) as it traverses a  weak stellar gravitational field, which we approximate using a polytropic model. We disregard the spin characteristic of the twisted particle, modeling it as a massless complex twisted scalar wave packet to simplify its interaction with gravitational fields. Building on this simplification, we determine the trajectory of this twisted particle by using the center of its energy density and investigate the gravitational birefringence induced by its OAM. In a weak field approximation, we find the gravitational birefringence-OAM relationship parallels that with spin, as described by the Mathisson-Papapetrou-Dixon equations. This indicates that the gravitational birefringence induced by OAM can potentially exceed that induced by spin by several orders of magnitude, significantly enhancing its detectability. To broaden our analysis, we introduce a nonminimal coupling term, $\lambda R|\phi|^2$, into the Lagrangian, resulting in the modified expression $\mathcal{L}=-\frac{1}{2}\nabla _\rho\phi\nabla^\rho\phi^*-\frac{1}{2}\lambda R|\phi|^2$. This adjustment is necessitated by the quantization of the scalar field in curved spacetime. We then explore the effects of this term on the motion of the  twisted particle. Our findings show that the trajectory of the twisted particle under nonminimal coupling ($\lambda\neq 0$) differs from that in the minimal coupling scenario ($\lambda=0$). Specifically, for a positive nonminimal coupling constant $\lambda$, the trajectory of the twisted particle is expected to deviate away from the stellar center, compared to the minimal coupling scenario.
\end{abstract}
\vspace{2pc}
\maketitle

\section{Introduction}\label{sec.intro}

Undoubtedly, in the field of modern physics, General Relativity (GR) holds its place as the most widely accepted theory of gravity yet. In this theory of gravity, the dynamics of spinning particles are proposed to be accurately described by using the Mathisson-Papapetrou-Dixon (MPD) equations \cite{Mathisson1937,papapetrou51,Dixon:1970zza,Dixon:1970zz,puetzfeld2015equations}. And an intriguing phenomenon was first discovered by Corinaldesi and Papapetrou, who found that particles with different angular momentum would follow differing trajectories in a gravitational field \cite{Corinaldesi1951}. This phenomenon, due to its similarity to the birefringence of light observed in certain materials, is commonly referred to as gravitational birefringence. This suggests that the motion of particles carrying angular momentum can provide a new avenue for the experimental testing of GR.

This phenomenon, gravitational birefringence, has been extensively discussed by using different methods. Several of these studies have approached this phenomenon by solving the MPD equations \cite{PhysRevD.67.024005,PhysRevD.90.064035,PhysRevD.96.043517,Duval2019}. In contrast, other researches discuss this phenomenon  by appling the Wentzel-Kramers-Brillouin (WKB) approximation to the Maxwell equations in GR \cite{PhysRevD.86.024010,Oancea2020,PhysRevD.104.025006,PhysRevD.105.104061,Andersson_2023}. Moreover, this phenomenon has been further investigated through the lens of the Energy-Momentum Tensor (EMT) \cite{PhysRevD.105.104008}. In another study, this phenomenon for free-falling electrons in a static uniform gravitational field has also been explored~\cite{PhysRevD.109.044060}. However, the primary focus of these studies has been on spin angular momentum, with its associated gravitational birefringence being too small to be detected. 

It is noteworthy that particles such as electrons, photons, atoms, and neutrons can exist in twisted states,  which possess intrinsic orbital angular momentum (OAM) \cite{Ivanov:2022jzh}. These particles, when in such twisted states, are referred to as "twisted" particles, highlighting their intrinsic OAM. As an example, electrons and photons in such twisted states are termed twisted electrons \cite{Bagrov:1980ct,Bliokh:2007ec,Bliokh:2017uvr} and twisted photons \cite{PhysRevA.45.8185}, respectively. For twisted photons in the lab environment, the OAM can exceed $10^4\hbar$ per photon \cite{doi:10.1073/pnas.1616889113}, and even reach up to $10^6\hbar$ per photon \cite{zhang_zhao2021}. Furthermore, it has been reported that light emitted from a Kerr black hole can also carry OAM \cite{Tamburini:2011tk,Yang:2014kyr}. This OAM has been claimed to be detected through the analysis of Event Horizon Telescope data \cite{Tamburini:2019vrf,Tamburini:2021jok}. The dynamics of these twisted particles, such as scalar particles with intrinsic OAM, can potentially differ markedly from that of spin-polarized particles, like spin-polarized photons. In GR, the former is described by the covariant Klein-Gordon equation, while the latter is governed by the covariant Maxwell's equations. Additionally, for a photon, distinct phenomena can arise from its spin and intrinsic OAM in certain experiments \cite{PhysRevLett.88.053601,PhysRevLett.91.093602,GHAI2009123,10.1117/1.AP.3.3.034001,Zhang:23,KU2023108777,PhysRevA.99.063832}, suggesting that their effects might be not interchangeable in some cases. Therefore, we think it is necessary to utilize alternative methods beyond the MPD framework to investigate the motion of twisted particles.

In GR, massless particles are expected to follow a null geodesic. When such a particle is twisted, possessing an intrinsic OAM, its trajectory might deviate away from the null geodesic due to its intrinsic OAM, as suggested by the MPD equations. In this work, we introduce a simplification by disregarding the spin characteristic of the twisted particle and representing it as a massless twisted complex scalar wave packet $\phi$, thus facilitating the analysis of its interactions with gravitational fields. Building on this simplification, we investigate the behavior of a free-falling twisted particle in a stellar gravitational field. As mentioned before, the dynamics of this twisted scalar particle are determined by the covariant Klein-Gordon equation in GR. It has been demonstrated that to ensure finiteness of the EMT of a scalar field at one-loop order, the Lagrangian density must include a nonminimal coupling term $\lambda R|\phi|^2$; furthermore, the nonminimal coupling constant $\lambda$ is required to be  $1/6$ \cite{Freedman:1974gs,Freedman:1974ze}. Here $R$ denotes the Ricci scalar, and $|\phi|^2$ signifies the square of the modulus of $\phi$. Correspondingly, the dynamics of the twisted particle are governed by the modified Klein-Gordon equation $\nabla_\mu\nabla^\mu\phi-\lambda R\phi=0$, with $\lambda$ representing the nonminimal coupling constant. This nonminimal coupling term arises from quantum corrections \cite{Linde:1982zj,PhysRevD.53.6813} and is expected to significantly affect the dynamics of twisted particles when the Ricci scalar $R$ of the spacetime is large. Therefore, the nonminimal coupling term might influence the motion of the twisted particle and lead to novel phenomena. Consequently, one of the main purposes of this study is to investigate the effects of nonminimal coupling and seek out potential phenomena.

Notably, the nonminimal coupling term $\lambda R|\phi|^2$ vanishes when the twisted particle moves outside gravitational sources, as the Ricci scalar $R=0$. Therefore, this work concentrates on the motion of the twisted particle inside stars, which are ubiquitous gravitational sources in the universe. We exclude all non-gravitational couplings, considering them outside the scope of this investigation. In this work, we employ the center of the EMT to describe the motion of the twisted particle. However, the EMT has 16 components, each leading to different trajectories. Additionally, the gravitational birefringence given by the energy flux density is also affected by another effect, named the Geometric Spin Hall Effect (GSHE) for its purely geometric nature, while the gravitational birefringence resulting from the energy density remains unaffected \cite{PhysRevLett.103.100401,Wang:2022bhh,PhysRevD.105.104008}. Compared with other components of the EMT, this unaffected property by GSHE suggests that the energy density has an advantage in investigating  the couplings between the OAM and gravitational fields. Consequently, our focus in this paper is primarily on the energy density. We reiterate that the topic of this work is to investigate the motion of a twisted particle in a stellar gravitational field and seek out potential phenomena induced by the nonminimal coupling.

 In this paper, we have set natural units where $\hbar=c=1$ and adopted the metric signature $(-,+,+,+)$. The curvature tensor, defined in terms of the Christoffel symbols $\Gamma^\alpha_{\mu\nu}$, is expressed as $R^\alpha_{\ \mu\nu\beta}\equiv \partial_\beta\Gamma^\alpha_{\mu\nu}-\partial_\nu\Gamma^\alpha_{\mu \beta}+\Gamma^\gamma_{\mu\nu}\Gamma^\alpha_{\beta \gamma}-\Gamma^\gamma_{\mu \beta}\Gamma^\alpha_{\nu \gamma}$. Furthermore, the Ricci tensor and Ricci scalar are denoted by $R_{\mu\beta}\equiv R^\gamma_{\ \mu\gamma\beta}$ and $R\equiv g^{\mu\nu}R_{\mu\nu}$, respectively. Greek indices run over the four coordinate labels in a general coordinate system, and Latin indices $i,j,k,\dots$ are restricted to three spatial coordinate labels.

This paper is organized as follows: Section \ref{sec.FW} provides an overview of the methodology used to investigate the motion of a twisted particle in a gravitational field. This section also introduces the configuration of the physical system and the numerical method used to solve the equations of motion (EoM) for a free-falling twisted particle. Section \ref{sec.GB} calculates the center of energy density of the twisted particle to describe its motion and details the gravitational birefringence induced by its OAM. Section \ref{sec.NM} compares the trajectory of the twisted particle under minimal and nonminimal coupling scenarios and discusses the effects of nonminimal couplings. Section \ref{sec.discu} offers further discussion of our findings.

\section{Equation of motion of free-falling twisted particles}\label{sec.FW}

Twisted particles, distinguished by their intrinsic OAM, diverge from point-like particles by requiring wave packet descriptions due to their complex OAM properties \cite{Ivanov:2022jzh}. The typical twisted particle is described by a Bessel or Laguerre-Gaussian function characterized by a nontrivial phase factor $\exp(i l \varphi)$ where $l$ is the OAM carrying by this twisted particle. In focusing solely on the effects of intrinsic OAM, our analysis deliberately excludes spin characteristics. The simplest method to examine the influence of intrinsic OAM in gravitational fields is through a massless twisted scalar particle model, employing a massless complex scalar wave packet. This approach has streamlined discussions on the generation of twisted photons near Kerr black holes when ignoring their spin \cite{Yang:2014kyr}. For a massless complex scalar field, the simplest and most reasonable nonminimal coupling term is given by $\lambda R|\phi|^2$. In this case, the Lagrangian of a free-falling twisted particle described by using this scalar field becomes
\begin{equation}\label{lang}
\mathcal{L}=-\frac{1}{2}\nabla_\mu\phi\nabla^\mu\phi^*-\frac{1}{2}\lambda R|\phi|^2,
\end{equation} 
where $\phi^*$ denotes the complex conjugate of $\phi$. The EoM of the twisted particle, determined by the principle of stationary action, is given by
\begin{equation}\label{evoequ}
\nabla_\mu\nabla^\mu\phi-\lambda R\phi=0,
\end{equation}
and its associated EMT takes the following form:
\begin{equation}\label{emt}
T^{\mu\nu}=\frac{1}{2}\nabla^\mu\phi\nabla^\nu\phi^*+\frac{1}{2}\nabla^\mu\phi^*\nabla^\nu\phi-\frac{1}{2}g^{\mu\nu}\nabla^\rho\phi\nabla_\rho\phi^*+\lambda(G^{\mu\nu}-g^{\mu\nu}\nabla_\rho\nabla^\rho+\nabla^\mu\nabla^\nu)(\phi\phi^*),
\end{equation}
where $G^{\mu\nu}\equiv R^{\mu\nu}-\frac{1}{2}g^{\mu\nu}R$  represents the Einstein tensor.

Before delving into the specifics of this work, let us provide a brief overview of the methodology employed to investigate the motion of the twisted particle in a gravitational field. Initially, we configure the physical system by establishing specific initial conditions and selecting an appropriate gravitational background. Subsequently, we solve the EoM for the twisted particle, as represented by equation~\eqref{evoequ}, using the predetermined initial conditions and gravitational background, a critical step that informs our subsequent discussion on the dynamics of the twisted particle. And then, we determine the trajectory of the twisted particle based on the center of its energy density:
\begin{equation}\label{center}
\langle x^i\rangle=\frac{\int^{+\infty}_{-\infty} \sqrt{g} x^i T^{00}dx^3}{\int^{+\infty}_{-\infty}\sqrt{g} T^{00}dx^3},
\end{equation}
where $g\equiv - \text{Det}(g_{\mu\nu})$ is the determinant of the metric tensor and $T^{00}$ is the energy density component of the EMT $T^{\mu\nu}$. Finally, we explore the gravitational birefringence and nonminimal coupling effects by examining the relationships between the trajectory, the OAM $l$, and the nonminimal coupling constant $\lambda$ of this twisted particle.

\subsection{The Physical System and Initial Conditions}

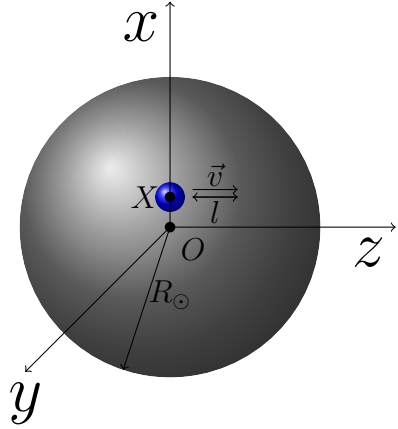
\begin{figure}
\centering
\usetikzlibrary {3d}
\begin{tikzpicture}[->,scale = 2]
\shade[ball color=gray,fill opacity=0.5] (0,0) circle (1.0);
\fill (0,0) circle (1pt) node[anchor=north west]{ $O$};
\shade[ball color=blue,fill opacity=0.8] (0,0.2) circle (0.1);
\draw (0.15,0.25,0) -> (0.45,0.25,0);
\node at(0.3,0.35,0){$\vec{v}$};
\draw (0,0,0) -> (-0.312,-0.95,0);
\node at(0,-0.45){$R_\odot$};
\draw (0.15,0.2,0) -> (0.45,0.2,0);
\draw (0.45,0.2,0)->(0.15,0.2,0) ;
\node at(0.3,0.1,0){$l$};
\fill (0,0.2) circle (1pt) node[anchor= east]{$X$};
\draw (0,0,0) -- (xyz cylindrical cs:radius=1.5) node[anchor=north east]{\huge $z$};%
\draw (0,0,0) -- (xyz cylindrical cs:radius=1.5,angle=90) node[anchor=north east]{\huge $x$};
\draw (0,0,0) -- (xyz cylindrical cs:z=2.5) node[anchor=north]{\huge $y$};
\end{tikzpicture}
 \caption{Schematic illustration of the physical system considered in this work. A twisted particle, shown as a blue sphere and carrying an OAM $l$, free-falls inside a star. The star, shown as a gray sphere, has a radius of $R_\odot$.}\label{sysfig}
\end{figure}

 As shown in figure~\ref{sysfig}, a twisted particle with an OAM $l$ free-falls inside a star, shown as a grey spherical region. Its initial position is located on the $x$-axis, and its initial velocity is directed along the $z$-axis. Throughout this paper, we set the  radius of the star to  $R_\odot=1$ facilitating the convenience of numerical computations.

As mentioned in section~\ref{sec.intro}, the nonminimal coupling term $\lambda R|\phi|^2$  is activated only when the twisted particle is within the gravitational source. Hence, it is preferable for the particle to be emitted from within the star, at an initial radial distance less than $R_\odot=1$. As shown in figure~\ref{gr-fr}, the gravitational potential reaches its maximum at a radial distance of $r\sim 0.4$ for the scenarios considered in this study. It is  reasonable to assume that a particle emitted from a location near this maximum potential point would be ideal to investigate the effects of intrinsic OAM and the nonminimal coupling. Moreover, it stands to reason that the longer a particle remains in the gravitational source, the more pronounced the effect of the nonminimal coupling term $\lambda R|\phi|^2$ becomes.

Given these considerations, it is reasonable to select the initial radial distance of the twisted particle near the peak of gravitational potential, such as $r_0=0.3$, $r_0=0.5$ or values in proximity to the maximum point at $r\sim 0.4$. This study aims to investigate the effects of intrinsic OAM and the non-minimal coupling term $\lambda R|\phi|^2$ on particle trajectories, alongside assessing the practicability of quantifying these influences. It is unlikely that the aforementioned selections of the initial position significantly alter these effects by orders of magnitude. Therefore, for our analysis, we chose $r_0=0.2$, a position that strikes a balance between being sufficiently close to both the stellar center and the point of maximum gravitational potential. This choice allows us to define the initial position in three-dimensional space as $\vec{X}=(0.2,0,0)$ at time $t=0$.

In the following part of this subsection, we aim to specify the initial conditions required to solve the EoM as outlined in equation~\eqref{evoequ}. Our objective is to define a set of initial conditions capable of determining a twisted scalar wave packet, which describes the free fall of the aforementioned twisted particle in a gravitational field. The methodology for identifying these initial conditions is as follows:
\begin{enumerate}
  \item \textbf{Construction of a Twisted Scalar Wave Packet}: We begin by constructing a twisted scalar wave packet, $\bar{\phi}(\vec{x}',t)$, in flat spacetime. This wave packet describes the free motion of a twisted particle.
  \item \textbf{Determination of Initial Conditions at $t=0$}: We then evaluate this wave packet and its time derivative at $t=0$. This process enables us to establish a set of initial conditions, $\bar{\phi}(\vec{x}',t)|_{t=0}$ and $\partial_t\bar{\phi}(\vec{x}',t)|_{t=0}$.
\end{enumerate}

The set of initial conditions $\bar{\phi}(\vec{x}',t)|_{t=0}$ and $\partial_t\bar{\phi}(\vec{x}',t)|_{t=0}$, representes the twisted wave packet's distribution in three-dimensional space at the initial time $t=0$. By integrating these conditions with the Klein-Gordon equation in flat spacetime, we can derive the twisted scalar wave packet $\bar{\phi}(\vec{x}',t)$. This packet models the free evolution of a twisted particle. Consequently, when these initial conditions are applied to the EoM~\eqref{evoequ}, they are expected to yield a twisted scalar wave packet that precisely represents the free-fall dynamics of the twisted particle in a gravitational field.

In a flat spacetime, the twisted particle possessing an OAM $l$, described by the scalar wave packet, can be represented using the Fourier Transform as
\begin{equation}\label{fourt}
\bar{\phi}(\vec{x},t)=\frac{1}{(2\pi)^{3/2}}\int^{+\infty}_{-\infty} \exp(-\frac{(\vec{k}-\vec{k}_0)^2}{2\Delta k^2}+il\varphi)\exp(i\vec{k}\cdot\vec{x}-i\omega t)dk^3,
\end{equation}
where $\omega=|\vec{k}|$ denotes the angular frequency and $\varphi$ is azimuthal angle of $\vec{k}$. In equation~\eqref{fourt}, the initial average momentum of the twisted particle ought to be $\vec{k}_0=(0,0,|\vec{k}_0|)$, considering that its initial velocity $\vec{v}$ is directed along the $z$-axis. 

For a specific wave packet, its wavelength $\lambda_\phi$ is typically far smaller than the size of this wave packet $d$. This characteristic implies that the magnitude of its initial average momentum, denoted as $k_0\equiv |\vec{k}_0|$, is significantly larger than the dispersion of its momentum $\Delta k$. The presence of the Gaussian factor $\exp\left(-\frac{(\vec{k}-\vec{k}_0)^2}{2\Delta k^2}\right)$ ensures a rapid decay of the function within the integral as $|\vec{k}-\vec{k}_0|/\Delta k$ increases. This indicates that $|\vec{k}-\vec{k}_0|\sim \Delta k$ is also substantially smaller than $|\vec{k}_0|$. Retaining terms to the second order of $|\vec{k}-\vec{k}_0|/k_0$, the angular frequency $\omega$ approximates to 
\begin{equation}\label{freq}
\omega\simeq k_0 + \delta k_\Vert +\frac{\delta k^2_\perp }{2k_0},
\end{equation}
where $\delta k_\Vert\equiv |\vec{k}\cdot\vec{k}_0-k_0^2|/k_0$ and $\delta k_{\perp}\equiv |(\vec{k}-\vec{k}_0)\times \vec{k}_0|/k_0$ represent the magnitudes of longitudinal and transverse components of $\vec{k}-\vec{k}_0$ with respect to $\vec{k}_0$.

By substituting equation~\eqref{freq} into equation~\eqref{fourt}, the integration can be carried out, yielding the scalar wave function $\bar{\phi}(\vec{x},t)$ as expressed by
\begin{equation}\label{wavef}
\begin{split}
\bar{\phi}(\vec{x},t)&\simeq 
\frac{\sqrt{\pi} r_\perp\Delta k^4 k_0^\frac{3}{2}}{2\sqrt{2}(k_0+it\Delta k^2)^\frac{3}{2}}\exp\left(-\frac{r_\perp^2\Delta k^2 k_0^2}{4(k^2_0+t^2\Delta k^4)}-\frac{(z-t)^2\Delta k^2}{2}\right)\\
&\times \left\{I_\frac{|l|-1}{2}\left(\frac{r_\perp^2\Delta k^2 k_0}{4(k_0+it\Delta k^2)}\right)-I_\frac{|l|+1}{2}\left(\frac{r_\perp^2\Delta k^2 k_0}{4(k_0+it\Delta k^2)}\right)\right\}\\
&\times \exp\left(i\vec{k}_0\cdot \vec{x}-ik_0 t+il\varphi+it\frac{r_\perp^2\Delta k^4 k_0 }{4(k^2_0+t\Delta k^4)}\right),\\
\end{split}
\end{equation}
where $r_\perp \equiv \sqrt{x^2+y^2}$ and $I_a(x)$ denotes the modified Bessel function of the first kind. The wave function $\bar{\phi}(\vec{x},t)$ is an approximate solution to the massless  Klein-Gordan equation in the flat spacetime $\partial_\mu\partial^\mu\bar{\phi}(\vec{x},t)=0$ with the following initial conditions:
\begin{equation}\label{initc}
\begin{split}
\bar{\phi}(\vec{x},t)|_{t=0}&= 
\frac{i^{|l|}\sqrt{\pi} r_\perp\Delta k^4}{2\sqrt{2}}\exp\left(-\frac{r^2_\perp\Delta k^2}{4}-\frac{z^2\Delta k^2}{2}\right)\\
&\times \left\{I_\frac{|l|-1}{2}\left(\frac{r^2_\perp\Delta k^2}{4}\right)-I_\frac{|l|+1}{2}\left(\frac{r^2_\perp\Delta k^2}{4}\right)\right\}\exp\left(i\vec{k}_0\cdot \vec{x}+il\varphi\right),\\
\partial_t\bar{\phi}(\vec{x},t)|_{t=0} &= -\frac{i^{|l|+1}\sqrt{\pi} \Delta k^4 r_\perp}{4\sqrt{2} k_0} \left\{ c_1 I_{\frac{|l|+3}{2}}\left(\frac{\Delta k^2 r^2_\perp}{4}\right)+\frac{c_2}{\Delta k^2 r^2_\perp}I_\frac{|l|+1}{2}\left(\frac{\Delta k^2 r^2_\perp}{4}\right)\right\} \\
&\times\exp \left(-\frac{\Delta k^2 \left(r^2_\perp+2 z^2\right)}{4}+i\vec{k}_0\cdot \vec{x}+il\varphi\right), \\
\end{split}
\end{equation}
where the two factors $c_1$ and $c_2$ are represented as
\begin{equation}
\begin{split}
c_1&=2 k_0^2-\Delta k^4 r^2_\perp+(2 i k_0 z+|l|+2)\Delta k^2, \\
c_2&=8 k_0^2 (1 +|l|) + \Delta k^6 r^4_\perp - \Delta k^4 r^2_\perp (6 + 3 |l| + 2 ik_0 z)\\
   & + 2 \Delta k^2 \left\{4 + 6 |l| + 2 |l|^2 - k_0^2 r^2_\perp + 4 ik_0 (1 + |l|) z\right\}.\\
 \end{split}
\end{equation}

 For analytical convenience, the center of the wave packet $\bar{\phi}(\vec{x},t)$ is initially located at the origin of the coordinate system at time $t=0$. To ensure consistency with the physical scenario in which the twisted particle is initially positioned at $\vec{X}=(0.2,0,0)$ at $t=0$, it becomes necessary to shift the wave packet $\phi(\vec{x},t)$ along the $x$-axis. It is necessary to adjust the initial conditions. Specifically, we redefine $r_\perp$ as $r'_\perp\equiv \sqrt{(x-0.2)^2+y^2}$ and $\vec{x}$ as $\vec{x}'\equiv\vec{x}-\vec{X}$ within equation~\eqref{initc}.  These modifications ensure that the initial conditions, $\bar{\phi}(\vec{x}',t)|_{t=}0$ and $\partial_t\bar{\phi}(\vec{x}',t)|_{t=0}$, describe a twisted wave packet centered at $\vec{X}$ at $t=0$. Consequently, these revised initial conditions are expected to determine a twisted scalar wave packet $\phi$, when they applied to the EoM \ref{evoequ}. This wave packet $\phi$ is expected to describe the free fall of a twisted particle starting from the initial position $\vec{X}$ within a gravitational field, as previously mentioned.

\subsection{The Gravitational Background}

 Despite excluding all non-gravitational couplings, it remains a considerable challenge to accurately simulate the distribution of matter inside a star, along with the corresponding gravitational field. However, our focus in this study is not on the internal structure of the star, but on the motion of a twisted particle. As such, we approximate the matter distribution within the star using a polytrope. This simplification allows for a straightforward calculation of the gravitational field.

 A polytrope is defined by its equation of state satisfying the given formula:
\begin{equation}\label{stat}
p(r)=K\rho(r)^\gamma,
\end{equation}
where $p(r)$ and $\rho(r)$ are the pressure and rest-mass density, and $r$ denotes the radial distance. Here, the constant $K$ is dependent on the entropy per nucleon and chemical composition in the star. In the polytrope model characterized by a specific $\gamma$, the constant $K$ for a given star can be determined by its mass $M$ and radius $R_\odot$. The process of solving the matter distribution of a polytrope and its corresponding internal gravitational field is a standard topic extensively covered in many textbooks on GR and stellar astrophysics. In the following discussion, we merely provide a brief overview. For a more detailed discussion, please refer to Chapter 11 in \cite{Weinberg:1972kfs}. Additionally, for those seeking to deepen their understanding of polytropes and their applications in astrophysics, the book referenced in \cite{2004ASSL306H} serves as an excellent introductory resource.

We introduce two variables, $\xi$ and $\theta(\xi)$, defined as follows:
\begin{equation}\label{vari}
r\equiv\left(\frac{K\gamma}{4\pi G(\gamma-1)}\right)^\frac{1}{2}\rho(0)^{(\gamma-2)/2}\xi,\quad \rho\equiv\rho(0)\theta^{1/(\gamma-1)}(\xi),
\end{equation}
where $\rho(0)$ is the rest-mass density at the center of this star. Consequently, the rest-mass density $\rho(r)$ can be determined by solving the Lane–Emden equation:
\begin{equation}\label{varequ}
\frac{1}{\xi^2}\frac{d}{d\xi}\left(\xi^2\frac{d\theta(\xi)}{d\xi}\right)+\theta^{1/(\gamma-1)}(\xi)=0
\end{equation}
with the following boundary conditions:
\begin{equation}\label{boun}
\theta(0)=0\qquad \theta'(0)=0,
\end{equation}
where $\theta'\equiv \frac{d\theta(\xi)}{d\xi}$. The radius of this star, denoted as $R_\odot$, can be obtained as:
\begin{equation}\label{radi}
R_\odot= \left(\frac{K\gamma}{4\pi G(\gamma-1)}\right)^\frac{1}{2}\rho(0)^{(\gamma-2)/2}\xi_1,
\end{equation}
where $\xi_1$ is the corresponding value of $\xi$ when $\rho=0$. 

\begin{figure}
\centering
 \subfigure{
 \includegraphics[scale=0.6]{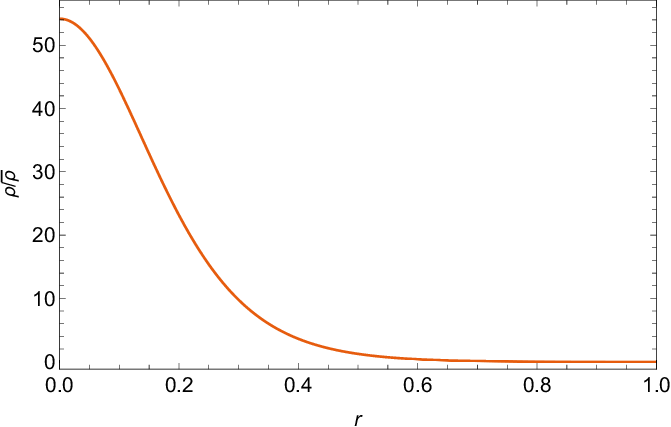}}
\subfigure{
 \includegraphics[scale=0.62]{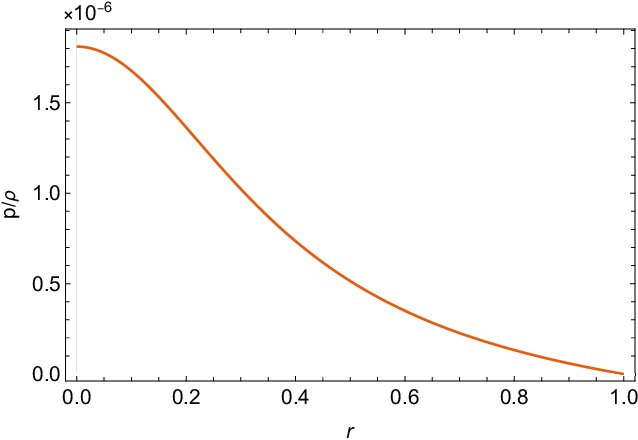}}
 \caption{The rest-mass density and pressure of a polytrope when $\gamma=4/3$. Here the radius of this polytrope has been set to $R_\odot=1$ and $\bar{\rho}$ is the average mass density.}\label{rho-p}
\end{figure}

Upon solving equation~\eqref{varequ} and considering the definition of $\theta(\xi)$ from equation~\eqref{vari} , we can determine the rest-mass density $\rho(r)$ and pressure $p(r)$ inside this star. As shown in figure~\ref{rho-p}, the ratio of pressure $p(r)$ to rest-mass density $\rho(r)$ is approximately $p(r)/\rho(r)\sim 10^{-6}\ll 1$. This indicates that the pressure $p(r)$ is greatly overshadowed by the rest-mass density $\rho(r)$ inside this star. 

\begin{figure}
\centering
 \subfigure{
 \includegraphics[scale=0.75]{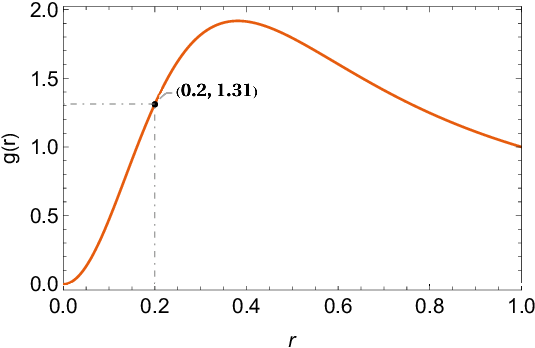}}
\subfigure{
 \includegraphics[scale=0.75]{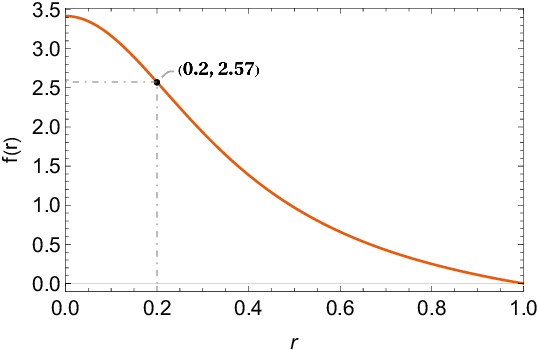}}
 \caption{The distributions of $g(r)$ and $f(r)$ for a polytrope when $\gamma=4/3$. Here the radius of this polytrope has been set to $R_\odot=1$.}\label{gr-fr}
\end{figure}

For a specific value of $\gamma$, the internal metric $g_{\mu\nu}$ of a star is determined by its surface gravitational potential, denoted as $\varepsilon\equiv \frac{GM_\odot}{R_\odot}$. Here, $M_\odot$ represents the total mass of the star. In this study, we concentrate on the case where $\varepsilon\ll 1$, indicating a weak gravitational field. This approximation is applicable to main-sequence stars, such as the Sun, for which $\varepsilon\sim 10^{-6}\ll 1$. In this case, the influence of pressure $p$ on the motion of particles within the star is expected to be significantly less, by approximately six orders of magnitude, than that of the mass density $\rho$. However, this influence of pressure is comparable to a second-order effect in terms of $\varepsilon$. Given this understanding, we elect to omit the consideration of pressure $p$ within the metric $g_{\mu\nu}$, directing our focus primarily towards the influence of mass density in this work.  By retaining only the first order of $\varepsilon$, the internal metric simplifies to
\begin{equation}\label{wemetr}
g_{\mu\nu}\simeq \eta_{\mu\nu}+h_{\mu\nu}, \quad h_{00}=2\varepsilon+2\varepsilon f(r), \quad h_{ij}=2\varepsilon \frac{g(r)x_i x_j}{r^2}, \quad h_{0i}=0
\end{equation}
with the following definitions:
\begin{equation}\label{defins}
  M(r)\equiv \int^r_0 4\pi r'^2\rho(r') dr',\quad  g(r)\equiv \frac{1}{\varepsilon}\frac{GM(r)}{r},\quad f(r)\equiv \frac{1}{\varepsilon}\int^{R_\odot}_{r}\frac{g(r')}{r'}dr',
\end{equation}
where $\eta_{\mu\nu}=\text{diag}(-1,+1,+1,+1)$ is the Minkowski metric. In this study, we adopt an ad hoc value of $\gamma=4/3$ for convenience. In accordance with equation~\eqref{defins}, we calculate the distributions of $g(r)$ and $f(r)$ for this polytrope using numerical methods, as displayed in figure~\ref{gr-fr}. At the initial position of the twisted particle where $r=0.2$, the gravitational potential is approximated by  $GM(r)/r\simeq 1.31\varepsilon$.
\subsection{Numerical Method}

 As demonstrated in equation~\eqref{wemetr}, the internal gravitational field of the star is weak, with the gravitational component in the metric $h_{\mu\nu}\propto \varepsilon \ll 1$. Retaining only the first order of $\varepsilon$, the corresponding affine connection $\Gamma^{\lambda}_{\rho\alpha}$ and Ricci scalar $R$ can be approximated as
\begin{equation}\label{conn}
\Gamma^{\lambda}_{\rho\alpha}\simeq\frac{1}{2}\eta^{\lambda\gamma}(\partial_\rho h_{\gamma\alpha}+\partial_\alpha h_{\gamma\rho}-\partial_{\gamma}h_{\rho\alpha}),\qquad
R \simeq -\partial_\lambda\partial_\alpha h^{\lambda\alpha},
\end{equation}
where $h^{\lambda\alpha}\equiv \eta^{\lambda\rho}\eta^{\alpha\beta}h_{\rho\beta}$. Therefore, equation~\eqref{evoequ} simplifies to
\begin{equation}
\eta^{\mu\nu}\partial_\mu\partial_\nu\phi =h^{\mu\nu}\partial_{\mu}\partial_{\nu}\phi+\partial_\alpha h^{\alpha\beta}\partial_\beta\phi-\frac{1}{2}\eta^{\alpha\beta}\partial_\alpha h\partial_\beta\phi-\lambda\partial_\alpha\partial_\beta h^{\alpha\beta}\phi,
\end{equation}
where $h\equiv \eta^{\mu\nu}h_{\mu\nu}$. In accordance with equation~\eqref{wemetr}, the EoM for the twisted particle inside the star can be expressed as
\begin{equation}\label{weom}
 \eta^{\mu\nu}\partial_\mu\partial_\nu\phi=2\varepsilon f(r)\eta^{ij}\partial_i\partial_j\phi +2\varepsilon\frac{g(r)x^i x^j}{r^2}\partial_i\partial_j\phi +3\varepsilon \rho x^i\partial_i\phi + 2\varepsilon g(r)\frac{x^i}{r^2}\partial_i\phi-6\lambda\varepsilon\rho'(r)\phi,
\end{equation}
where $\varepsilon\rho'(r)\equiv \rho(r)/\bar{\rho}$.

 It is nearly impossible to find an analytical solution for the EoM given by equation~\eqref{weom} with the initial conditions as represented by equation~\eqref{initc}. Consequently, we resort to numerical methods to solve the EoM in this work. It is worth noting that the wave function $\phi$ is expected to exhibit rapid oscillations in both space and time due to the term $\exp(i\vec{k}_0\cdot\vec{x}-ik_0 t)$, as suggested by equation~\eqref{wavef}. This implies that the time and space step sizes in the numerical calculations must be significantly smaller than the wavelength $\lambda_\phi=2\pi/k_0$. However, reducing the time and space step sizes increases the computational demand. Therefore, we must simplify equation~\eqref{weom} for efficient numerical computations.
 
 Let us introduce a new wave function $\phi'$, defined as
 \begin{equation}\label{phip}
 \phi\equiv \phi'\exp(i\vec{k}_0\cdot\vec{x}-ik_0 t).
 \end{equation}
The EoM for $\phi'$ can be derived from equation~\eqref{weom}, which yields
\begin{equation}\label{evophip}
\begin{split}
&\quad -\partial_t^2\phi'+2ik_0\partial_t\phi'+\eta^{ij}\partial_i\partial_j\phi'+2ik_0\delta_3^i\partial_i\phi'\\
&=2\varepsilon \left( f(r)\eta^{ij} + \frac{g(r)x^i x^j}{r^2}\right)\left(\partial_i\partial_j\phi'+ik_0\delta^3_i\partial_j\phi'+ik_0\delta^3_j\partial_i\phi'-k^2_0\delta^3_i\delta^3_j\phi' \right)\\
&+\varepsilon \left( 3\rho+\frac{2g(r)}{r^2} \right)x^i(\partial_i\phi'+ik_0\delta^3_i\phi')-6\lambda\varepsilon
\rho'(r)\phi',\\
\end{split}
\end{equation}
where the functions $g(r)$ and $f(r)$ have been defined in equation \eqref{defins}. As suggested by equation~\eqref{wavef}, the rapid oscillations in the wave function $\phi'$ are expected to vanish. Consequently, the spatial step size can significantly exceed the wavelength $\lambda_\phi$. 

Significantly, equation~\eqref{evophip} remains a second-order inhomogeneous partial differential equation in four-dimensional spacetime, the complexity of which poses substantial challenges to even numerical calculations. To simplify, we apply a Fourier Transform to this equation in three-dimensional space, thereby transforming equation~\eqref{evophip} into an ordinary differential equation. Consequently, the EoM of $\phi'$  simplifies to
  \begin{equation}\label{feom}
 \partial^2_t\phi'_f -2ik_0\partial_t\phi'_f - k_0^2\phi'_f= -k'_i k'^i\phi'_f + 2\varepsilon \mathcal{F}^{ij}*(k'_i k'_j \phi'_f) - i\varepsilon \mathcal{G}^i*(k'_i\phi_f)- 6\lambda \varepsilon \rho'_f*\phi'_f,
  \end{equation}
 given the following definitions:
 \begin{equation}
 \begin{split}
 &\vec{k}'\equiv (k^1,k^2,k^3+k_0),\quad \mathcal{F}^{ij}\equiv\mathscr{F}\left(f(r)\eta^{ij} + \frac{g(r)x^i x^j}{r^2}\right),\quad \mathcal{G}^i\equiv \mathscr{F}\left( 3\rho x^i+\frac{2g(r)}{r^2}x^i \right)\\
 &\rho'_f\equiv \mathscr{F}(\rho'),\quad \phi'_f\equiv \mathscr{F}(\phi'),\\
 \end{split}
 \end{equation}
where $\mathscr{F}(\phi)\equiv \frac{1}{(2\pi)^{3/2}}\int^{+\infty}_{-\infty} \phi(x)\exp(-i\vec{k}\cdot\vec{x})dx^3$ is the Fourier Transform in three-dimensional space and the operator '$*$' signifies the convolution in three-dimensional space $(f*g)(\vec{x})\equiv \int^{+\infty}_{-\infty} f(\vec{x}')g(\vec{x}-\vec{x}')dx'^3$. 

  There exists a plethora of numerical methods for solving this ordinary differential equation \eqref{feom}. In this study, we employ the midpoint method with the initial conditions previously mentioned. Once $\phi'_f$ is obtained  using this method, the wave function $\phi$ can be effortlessly calculated as
  \begin{equation}\label{phi}
  \phi(\vec{x},t)=\exp(i\vec{k}_0\cdot \vec{x}-ik_0 t)\mathscr{F}^{-1}(\phi'_f(\vec{k},t)),
  \end{equation}
where $\mathscr{F}^{-1}$ denotes the inverse Fourier Transform of $\mathscr{F}$. 

In this paper, unless otherwise specified, the numerical calculations are performed with the following default parameters: the wavelength $\lambda_\phi=\pi/2500$, the OAM $l=+1$, and the gravitational potential at the surface of the star $\varepsilon=10^{-3}$. Here, the positive sign in "$l=+1$" indicates that the intrinsic OAM is aligned with the momentum of the twisted particle.

\section{Gravitational Birefringence Induced By OAM}\label{sec.GB}

Numerous researchers have pointed out that the gravitational birefringence occurs in the direction perpendicular to both the gravity and angular momentum \cite{Duval2019,Oancea2020,PhysRevD.105.104008,PhysRevD.109.044060}. Given the  geometric configuration of the physical system shown in figure~\ref{sysfig}, this phenomenon manifests along the direction parallel to the $y$-axis. Therefore, this section primarily focuses on the trajectory along the $y$-axis under the minimal coupling condition $\lambda=0$.

\begin{figure}
\centering
 \subfigure{
 \includegraphics[scale=0.7]{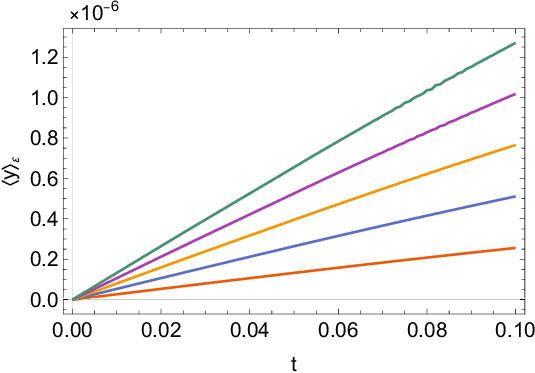}
 \includegraphics[scale=0.72]{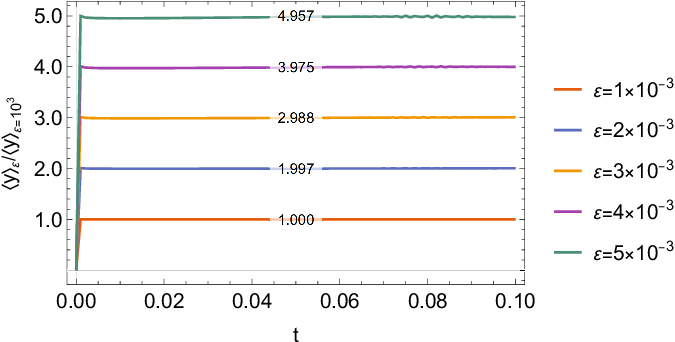}}\\
  \subfigure{
 \includegraphics[scale=0.72]{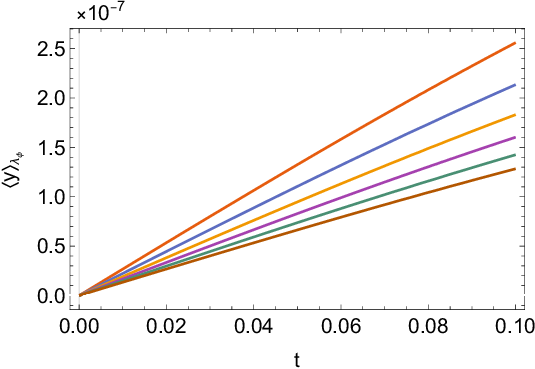}
 \includegraphics[scale=0.73]{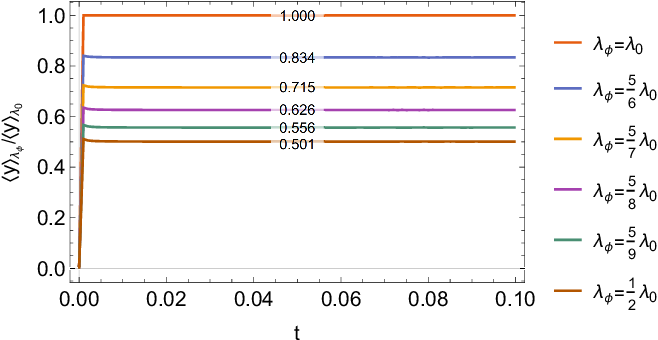}}\\
   \subfigure{
  \includegraphics[scale=0.72]{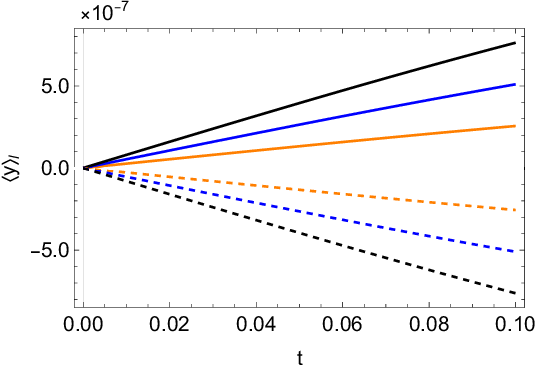}
 \includegraphics[scale=0.72]{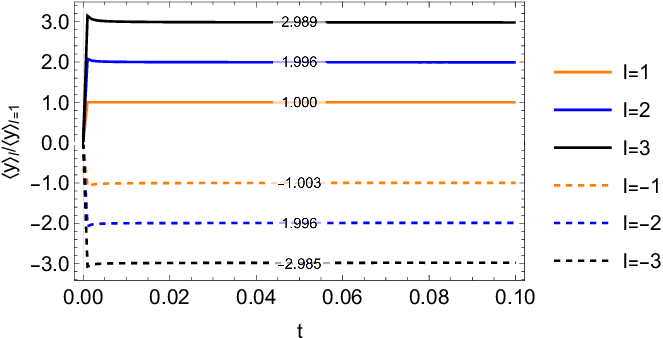}}
 \caption{The trajectories of twisted particles along the $y$-axis.}\label{grabi.fig}
\end{figure}

The trajectory along the $y$-axis, represented by the the temporal evolution of $\langle y\rangle$ is shown in figure~\ref{grabi.fig}. The analysis reveals an approximate linear dependence of $\langle y\rangle$ on the time $t$ for $t < 0.1$. It is also indicated that $\langle y\rangle$  is proportional to the wavelentgh $\lambda_\phi$, surface  gravitational potential of the star $\varepsilon$ and OAM $l$, respectively. Therefore, the trajectory along the $y$-axis can be approximately given by the following relation:
\begin{equation}\label{gboam}
\langle y\rangle \simeq 2.06\varepsilon l\lambda_\phi t.
\end{equation} 

According to \cite{Duval2019,PhysRevD.105.104008}, during a short propagation period $t\ll 1$, the trajectory of a spin-polarized photon along the $y$-axis is given by 
\begin{equation}
\langle y\rangle_\sigma \simeq\frac{GM}{b}\frac{\sigma\lambda_\gamma t}{\pi b},
\end{equation}
where $b$ denotes the impact parameter, and $GM/b$ represents the gravitational potential at the photon's initial position. The photon's wavelength is represented by $\lambda_\gamma$, and its helicity by $\sigma$ which takes the values of $\pm 1$. As mentioned in section~\ref{sec.FW}, we have set an impact parameter for the twisted scalar particle as $b=0.2$, and the corresponding gravitational potential $GM(r)/b\simeq 1.31\varepsilon$. Consequently, for a spin-polarized photon emitted from the same location, its trajectory along the $y$-axis can be described by the following relation:
\begin{equation}\label{gbspin}
\langle y\rangle_\sigma \simeq 2.08 \varepsilon \sigma\lambda_\gamma t.
\end{equation}
The form of this equation mirrors that of equation~\eqref{gboam}, with factor values in close agreement. Consequently, the gravitational birefringence-OAM relationship parallels that of spin, suggesting that OAM's effects on particle motion are analogous to those of spin.

\begin{figure}
  \centering
   \subfigure{
   \includegraphics[scale=0.8]{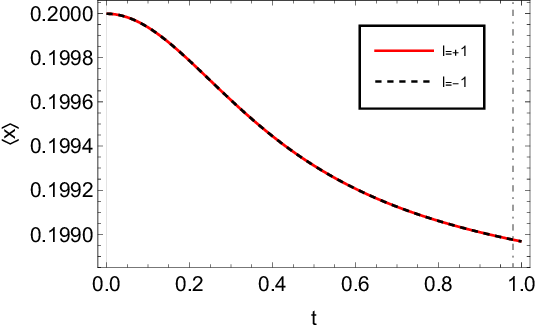}
   \includegraphics[scale=0.76]{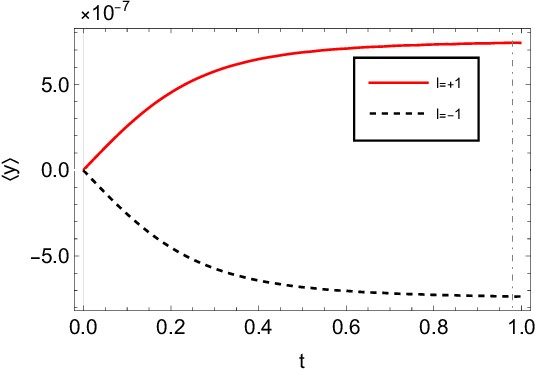}}\\
   \subfigure{
   \includegraphics[scale=0.78]{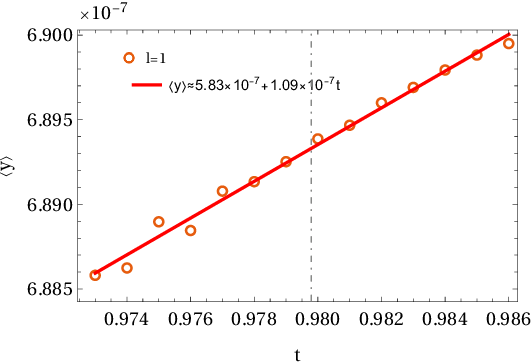}
   \includegraphics[scale=0.79]{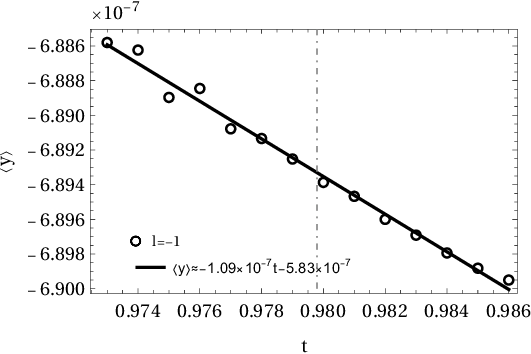}}\\
   \caption{The trajectories of two twisted particles with opposite OAM $l=\pm 1$ as they propagate from inside to outside of the star. The lower panel details their trajectories along the $y$-axis near the stellar surface, with the dot-dashed line representing the surface of the star.}\label{grabi-lt.fig}
  \end{figure}

During the movement of the twisted particle from inside to outside of the star, the wavelength $\lambda_{\phi}$, OAM $l$ and surface gravitational potential of the star $\varepsilon$, are expected to remain constant. Consequently, the proportionality $\langle y \rangle \propto \varepsilon \lambda_{\phi} l$ is anticipated to be preserved. However, due to significant spatial variations in the gravitational field during the movement, the trajectory along the $y$-axis will inevitably deviate from the simple linear function of time $t$. The numerical results are shown in figure~\ref{grabi-lt.fig}, which indicate that upon reaching the stellar surface, the velocity  along the $y$-axis of the twisted particle is approximately given by
\begin{equation}\label{y-vel}
\frac{d\langle y\rangle_l}{dt}\arrowvert_{surf}\simeq  1.09\times 10^{-7}l\simeq 0.08\varepsilon l \lambda_\phi,
\end{equation}
where the parameters $\lambda_\phi=\pi/2500$, $\varepsilon=10^{-3}$, $l=\pm 1$ have been considered, along with the previously established proportionality $\langle y\rangle \propto \varepsilon l\lambda_\phi$. Consequently, the angle $\theta\simeq \frac{2d\langle y\rangle}{dt}$,  representing the angular separation along the $y$-axis between two twisted particles with opposite OAM $l=\pm |l|$, is approximated as
\begin{equation}\label{y-ang}
\theta\sim 0.16\times \varepsilon |l|\lambda_\phi.
\end{equation}

It is worth noting that the angle $\theta$ is proportional  to both the wavelength $\lambda_\phi$ and OAM $l$. Therefore, an increase in OAM leads to a more pronounced gravitational birefringence. Under controlled laboratory conditions, it has been demonstrated that light can carry an OAM as high as $10^6\hbar$ per photon \cite{zhang_zhao2021}. As such, the gravitational birefringence induced by OAM could significantly exceed that induced by spin, potentially facilitating its detection. Similarly, an increase in wavelength also enhances the effect of gravitational birefringence. Therefore, detection using radio waves might prove easier than using light or gamma rays.

\section{Nonminimal Coupling Effects}\label{sec.NM}

 According to equations~\eqref{evoequ} and \eqref{emt}, variations in the nonminimal coupling constant $\lambda$ induce changes in both the wave function $\phi$ and EMT $T^{\mu\nu}$, irrespective of identical initial conditions. As a result, the energy density center of the twisted particle is expected to shift as $\lambda$ varies. Consequently, the trajectory of the twisted particle, defined by its energy density center, can alter with different values of $\lambda$, as the twisted particle free falls inside a gravitational source. The effects of nonminimal coupling on the motion of twisted particles may reveal a novel phenomenon. 
 
 Moreover, as suggested by equation~\eqref{evoequ}, the nonminimal coupling effects are expected to be proportional to the nonminimal coupling constant $\lambda$. In this section, we explore a large range of arbitrary values for $\lambda$ within the interval $0\leqslant\lambda\leqslant 10^5$ to investigate these potential effects. The discussion section will then revisit the specific effects of nonminimal coupling when $\lambda=1/6$, as initially introduced in section~\ref{sec.intro}.

\begin{figure}
\centering
 \subfigure{
 \includegraphics[scale=0.75]{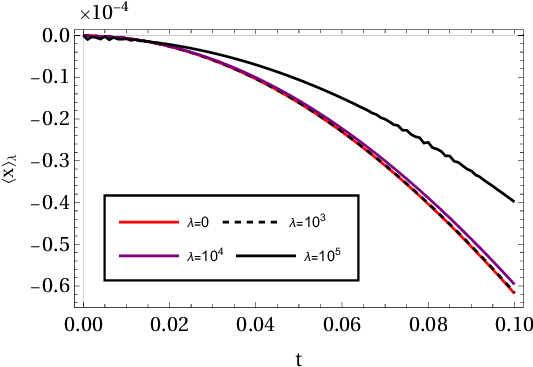}
   \hspace{5pt}
 \includegraphics[scale=0.75]{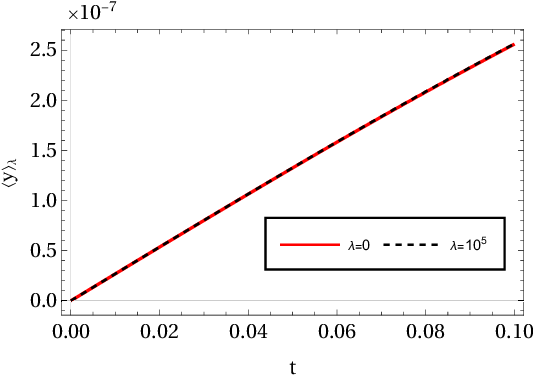}}
 \caption{The trajectory of the free-falling twisted particle with different nonminimal coupling constant $\lambda$.  In the left panel, the initial position of the twisted particle is aligned with the origin of the $x$-axis.}\label{gra-non.fig}
\end{figure}

After numerical calculations, the trajectory of the twisted particle with different nonminimal coupling constants $\lambda$ is shown in figure~\ref{gra-non.fig}. It is indicated that the nonminimal couplings can affect the trajectory along the $x$-axis, while leaving the trajectory along the $y$-axis negligibly affected. Therefore, in this section, we mainly focus on the trajectory along the $x$-axis which differs significantly with different $\lambda$. In the remainder of this section, unless otherwise specified, the numerical calculations are performed with the default nonminimal coupling constant $\lambda=10^4$.

\begin{figure}
\centering
  \subfigure{
 \includegraphics[scale=0.72]{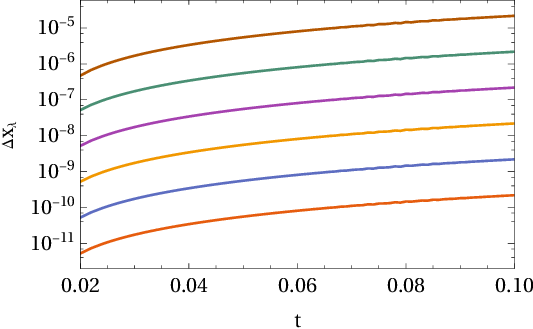}
 \includegraphics[scale=0.7]{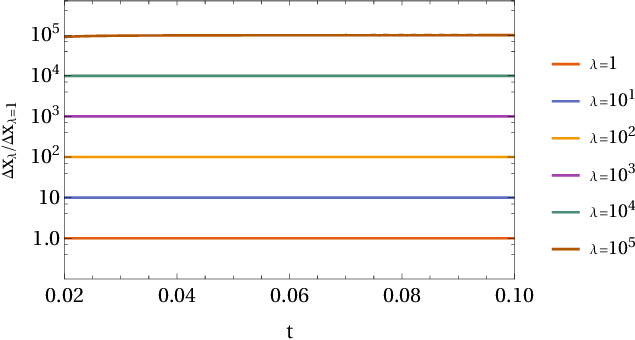}}\\
   \subfigure{
 \includegraphics[scale=0.72]{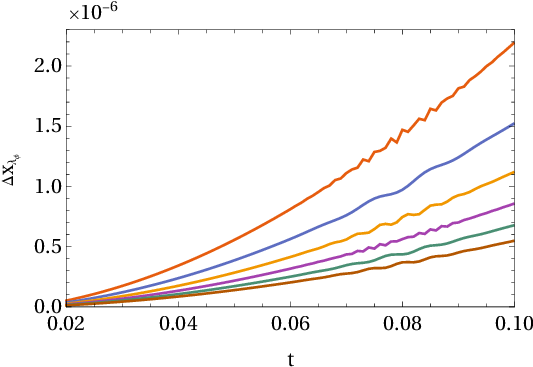}
 \includegraphics[scale=0.73]{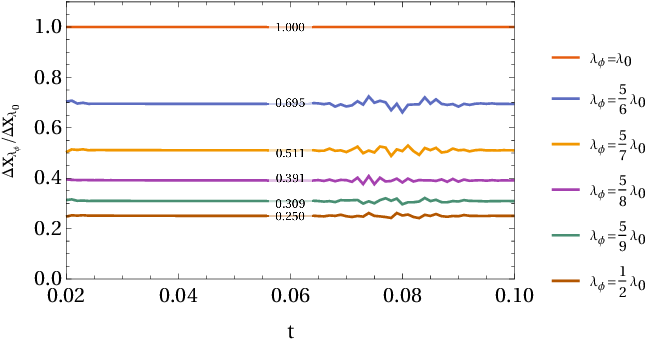}}\\
    \subfigure{
 \includegraphics[scale=0.65]{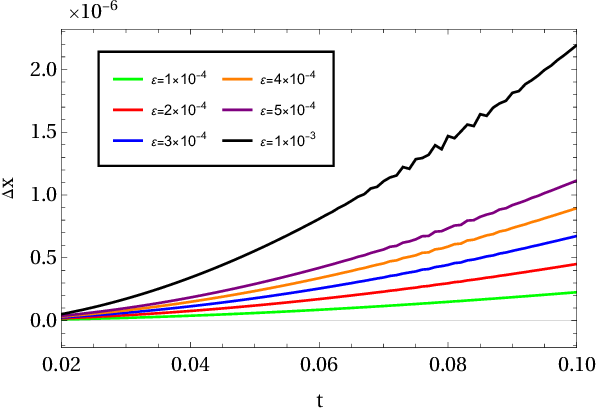}
 \includegraphics[scale=0.73]{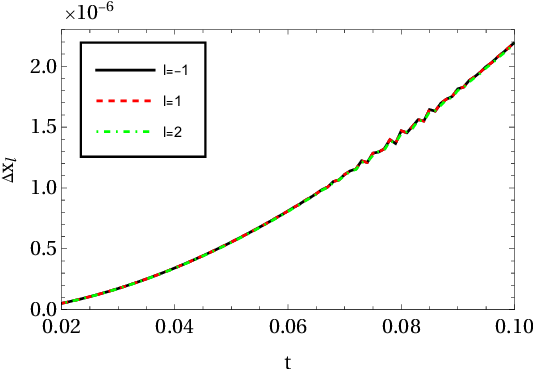}}\\
 \caption{The deviation of the trajectory along the $x$-axis between the nonminimal and minimal coupling cases.}\label{gradx-non.fig}
\end{figure}

The numerical results, presented in figure~\ref{gradx-non.fig}, indicate that the deviation of the trajectory along the $x$-axis, denoted as $\Delta x\equiv \langle x\rangle_\lambda-\langle x\rangle_{\lambda=0}$, between the scenarios of nonminimal and minimal couplings is approximately proportional to the nonminimal coupling constant $\lambda$, the surface gravitational potential of the star $\varepsilon$, and the square of the wavelength $\lambda_\phi^2$, respectively. Our numerical results further suggest that, in the regime where $\lambda_\phi/R_\odot \ll 1$, the influence of OAM $l$ on the trajectory deviation $\Delta x$ is negligible. Therefore, the relationship between the deviation $\Delta x$ and the aforementioned parameters can be approximately represented as
\begin{equation}\label{eff-non}
\Delta x\propto \varepsilon\lambda\lambda_\phi^2.
\end{equation}

\begin{figure}
\centering
 \subfigure{
 \includegraphics[scale=0.8]{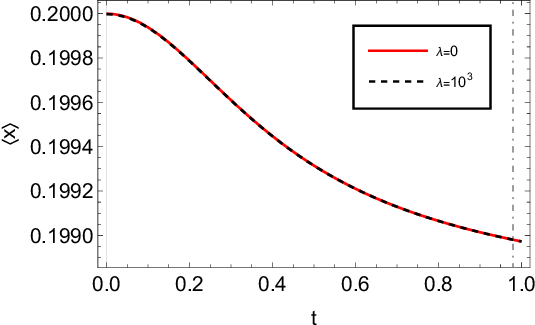}
   \hspace{5pt}
 \includegraphics[scale=0.75]{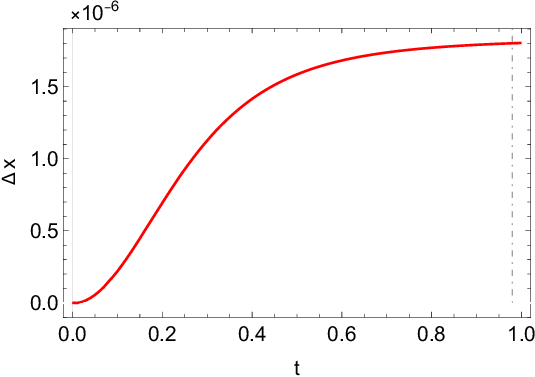}}
 \caption{The trajectory along the $x$-axis and its deviation as the twisted particle propagates from inside to outside of the star within the scenarios of nonminimal and minimal couplings.  The dot-dashed line representes the surface of the star.}\label{gradx-lt-non.fig}
\end{figure}

The trajectory of the twisted particle along the $x$-axis during its movement from inside to outside of the star is depicted in figure~\ref{gradx-lt-non.fig}. Its velocity along the $x$-axis, $v_x$, can be defined by $v_x\equiv \frac{d\langle x\rangle}{dt}$.  Correspondingly, the difference in $v_x$ between the nonminimal and minimal coupling scenarios is represented by
\begin{equation}\label{def-dvx}
\Delta v_x\equiv -\left(\frac{d\langle x\rangle_\lambda}{dt}-\frac{d\langle x\rangle_{\lambda=0}}{dt}\right) =-\frac{d\Delta x}{dt},
\end{equation}
where the negative sign indicates that, with a positive nonminimal coupling constant $\lambda$, the trajectory deviates along the direction away from the stellar center when compared to the trajectory in the minimal coupling case.

Numerical results shown in figure~\ref{gradx-lt-non.fig} indicate that, upon the twisted particle reaching the stellar surface, $\Delta v_x$ is approximately given by
\begin{equation}\label{dvx-lt}
\Delta v_x|_{surf}\simeq -1.08\times 10^{-7}.
\end{equation}
As the twisted particle moves from inside to outside of the star, the proportionality $\Delta x\propto \varepsilon\lambda\lambda_\phi^2$ is also anticipated to be preserved, mirroring the reasons underlying the previously established $\langle y \rangle \propto \varepsilon \lambda_{\phi} l$. Correspondingly, the difference in $v_x$ at the stellar surface, denoted as $\Delta v_x|_{surf}$, follows the established proportionality $\Delta v_x|_{surf}\propto -\varepsilon\lambda\lambda_\phi^2$. Given the parameter settings $\lambda=10^3$, $\varepsilon=10^{-3}$ and $\lambda_\phi=\pi/2500$ in the numerical calculations to obtain equation~\eqref{dvx-lt}, $\Delta v_x|_{surf}$ can be approximately represented as
\begin{equation}\label{dvx}
\Delta v_x|_{surf}\simeq -0.07\times \varepsilon\lambda\lambda_\phi^2.
\end{equation}

Outside the star, the Ricci scalar $R$ becomes zero, leading to the vanishing of the nonminimal coupling term $\lambda R|\phi|^2$. Consequently, the velocity difference $\Delta v_x$ for the twisted particle becomes constant upon exiting the stellar surface. Therefore, equation~\eqref{dvx} suggests that the trajectory in the case of nonminimal coupling deviates from the minimal coupling case by an angle $\theta_{non}$ along the $x$-axis, which can be approximated as
\begin{equation}\label{dangx}
\theta_{non}\sim -0.07\times \frac{\varepsilon\lambda\lambda_\phi^2}{R_\odot^2},
\end{equation} 
where the radius of the star $R_\odot$ explicitly appears to maintain the generality of the formula. As indicated by equation~\eqref{dangx}, the angle $\theta_{non}$ scales with $\lambda_\phi^2/R_\odot^2$. Given that the wavelength $\lambda_\phi$  of a realistic wave packet is typically at least several orders of magnitude smaller than the stellar radius $R_\odot$, the angle $\theta_{non}$ induced by the nonminimal coupling term $\lambda R|\phi|^2$ is expected to be drastically suppressed.

\section{Discussion}\label{sec.discu}

As previously established, the trajectories of two free-falling twisted particles carrying opposite OAM exhibit a transverse  separation  perpendicular to both the OAM and gravity, which is a manifestation of gravitational birefringence. This separation obeys the proportionality $\theta\propto \varepsilon l\lambda_\phi/R_\odot$, consistent with the well-documented relationship between the separation and spin angular momentum. This consistency implies that, when coupled to gravity, the effects induced by OAM are analogous to those induced by spin, as described by MPD equations.

For a twisted particle, the OAM can significantly surpass its intrinsic spin, potentially by orders of magnitude. This substantial difference suggests that the gravitational birefringence induced by OAM might be much larger than that induced by spin. To illustrate, consider the birefringence of sunlight  at a typical wavelength of $\lambda_\phi\sim 500\text{nm}$ as it passes through the gravitational field of the Sun. The spin of light is limited to $\hbar$ per photon, while its OAM can exceed $10^4\hbar$ per photon, as reported in \cite{doi:10.1073/pnas.1616889113}. The surface gravitational potential of the Sun is approximated to be $\varepsilon\sim 2.43\times 10^{-6}$, with a radius of $R_\odot\sim 6.96\times 10^{5}\text{km}$. Consequently, the  separation induced by the spin of light is estimated to be $\theta\sim 10^{-22}\text{rad}$, whereas the separation due to its OAM is expected to be $\theta\sim 10^{-18}\text{rad}$,  which is an increase by four orders of magnitude over that induced by the spin. This substantial amplification associated with OAM could significantly enhance the detectability of gravitational birefringence, thereby improving the prospects for experimental verification.
 
The numerical results suggest that a free-falling twisted particle is expected to follow different trajectories with variations in the nonminimal coupling constant $\lambda$. Specifically, as the twisted particle propagates from inside to outside of a star along a non-radial direction, its trajectory in the case of nonminimal coupling deviates from the minimal coupling case by an angle $\theta_{non}\sim -0.07\times \varepsilon\lambda\lambda_\phi^2/R_\odot^2$ along the direction of gravity. The negative sign here signifies that, for a positive coupling constant $\lambda$, the deviation is directed away from the stellar center.

Due to the proportionality  of $\lambda_\phi^2/R_\odot^2$, this angle $\theta_{non}$ is expected to be many orders of magnitude smaller than the deflection angle $\Delta \varphi$ predicted by GR, which can be approximated by $\Delta \varphi\sim \varepsilon$. For a twisted particle emitted from the inside of the Sun with a typical wavelength $\lambda_\phi\sim 500\text{nm}$, the ratio of its wavelength to the radius of the Sun  is approximated as $\lambda_\phi/R_\odot\sim 7.18\times 10^{-16}$. In accordance with the discussion in section~\ref{sec.intro}, for $\lambda=1/6$,  the estimated nonminimal coupling angle $\theta_{non}$ is calculated to be approximately $ -10^{-38}\text{rad}$. This value is $32$ orders of magnitude smaller than the gravitational deflection angle $\Delta \varphi$, which is on the order of $10^{-6}\text{rad}$. With current technology, detecting an angle as small as $\theta_{non}$ is almost impossible.

It is important to note that both the angles, $\theta$ and $\theta_{non}$, are proportional to the gravitational potential $\varepsilon$. The angle $\theta$ is inversely proportional to the radius of the gravitational source $R_\odot$, while the angle $\theta_{non}$ is inversely proportional to its square, $R_\odot^2$. Therefore, with an increase in the gravitational potential $\varepsilon$ and a decrease in the radius of its source, the angles $\theta$ and $\theta_{non}$ are expected to increase significantly. Although these results were obtained under the weak field approximation, where the gravitational potential $\varepsilon\ll 1$, using them for order-of-magnitude estimates of these effects near compact objects remains reasonably justified. As an example, replacing the Sun with a pulsar having a gravitational potential $\varepsilon \sim 0.1$ at a radius of $R_{\odot} \sim 10$ km, the angle $\theta$ with the OAM $l=10^4$ is expected to become $\theta \sim  10^{-8}\text{rad}$, which is nearly $10$ orders of magnitude greater than that in the gravitational field of the Sun. Meanwhile, the angle $\theta_{non}$ is expected to be $\theta_{non} \sim -10^{-24}\text{rad}$, nearly $14$ orders of magnitude larger than that in the gravitational field of the Sun. These results indicate that, in the gravitational fields of compact objects such as black holes and neutron stars, the angles $\theta$ and $\theta_{non}$ could potentially be large enough to be detectable.

The trajectory of a twisted particle is likely influenced by its form of nonminimal coupling to gravity, which we have modeled as $\lambda R|\phi|^2$ in this work. Changing this coupling form could modify the nonminimal coupling outcomes. Our findings confirm that particle trajectories are sensitive to their nonminimal couplings, indicating the potential to test gravitational theories by comparing trajectories under different coupling forms.

\ack{We are grateful to Xiang-Song Chen for the helpful discussions, and we thank Zheng-Cheng Liang for his suggestions during the numerical calculations. This work is  supported by the National Natural Science Foundation of China(Grant No. 12375084).}	

\appendix
\section{The twisted scalar wave packet $\bar{\phi}(\vec{x},t)$ in the flat spacetime}

In the flat spacetime, the dynamics of a massless complex scalar field are described by the Klein-Gordan equation:
\begin{equation}\label{K-G}
  \eta^{\rho\alpha}\partial_\rho\partial_\alpha \phi = 0,
\end{equation}
where $\eta^{\rho\alpha}=\text{diag}(-+++)$ represents the Minkowski metric. This equation admits a plane-wave solution of the form:
\begin{equation}
  \phi_p=\exp(i\vec{k}\cdot{x}-i\omega t),
\end{equation}
with $\omega=|\vec{k}|$. By using these plane waves, one can construct a wave packet $\bar{\phi}(\vec{x},t)$ as follows:
\begin{equation}
  \bar{\phi}(\vec{x},t)=\frac{1}{(2\pi)^{3/2}}\int^{+\infty}_{-\infty} \exp\left(-\frac{(\vec{k}-\vec{k}_0)^2}{2\Delta k^2}+il\varphi\right)\exp(i\vec{k}\cdot\vec{x}-i\omega t)dk^3,
\end{equation}
where $\varphi$ denotes the azimuthal angle of $\vec{k}$, and $\Delta k$ representes the dispersion of this wave packet in momentum space. It can be verified that the wave packet $\bar{\phi}(\vec{x},t)$ not only satisfies the Klein-Gordan equation but also carries an intrinsic orbital angular momentum (OAM) of $l$. For analytical convenience, the wave packet $\bar{\phi}(\vec{x},t)$ can be reformulated as:
\begin{equation}\label{app1}
  \bar{\phi}(\vec{x},t)=\frac{1}{(2\pi)^{3/2}}\exp(i\vec{k}_0\cdot\vec{x})\int^{+\infty}_{-\infty} \exp\left(-\frac{\delta k^2}{2\Delta k^2}+il\varphi\right)\exp(i\delta\vec{k}\cdot\vec{x}-i\omega t)d\delta k^3
\end{equation}
with the following definitions:
\begin{equation}
  \delta \vec{k}=\vec{k}-\vec{k}_0,\quad \delta k=|\delta\vec{k}|, \quad \omega = |\vec{k}_0 + \delta \vec{k}|.
\end{equation}

In the case of a realistic wave packet, the wavelength $\lambda$ is typically much smaller than the size of this wave packet $d$. This characteristic is crucial for the wave packet's longevity, as a comparable size would lead to rapid dispersion and short-lived existence. The wave packet $\bar{\phi}(\vec{x},t)$ incorporates a Gaussian distribution in momentum space, expressed as $\exp(-\frac{\delta k^2}{2\Delta k^2})$. This distribution implies that the wave packet's wavelength can be approximated by $\lambda\sim 1/|\vec{k}_0|$, and its spatial radius by $d\sim 1/\Delta k$. Consequently, the condition $\Delta k\ll |\vec{k}_0|$ arises from the inequality $\lambda\ll d$. Additionally, within the bounds of integration in equation \eqref{app1}, it is reasonable to assume $\delta k\sim \Delta k$ given the rapid decrease of the Gaussian factor with increasing $\delta k/\Delta k$. This assumption leads to the conclusion $\delta k\ll |\vec{k}_0|$. 

Given these considerations, the frequency $\omega$ can be expanded in terms of $\delta k/|\vec{k}_0|$up to the second order, yielding:
\begin{equation}
  \omega\simeq k_0 + \delta k_\Vert +\frac{\delta k^2_\perp }{2k_0}
  \end{equation}
with $k_0=|\vec{k}_0|$. Here, $\delta k_\Vert=\delta\vec{k}\cdot\vec{k}_0/k_0$ and $\delta k_{\perp}= |\delta\vec{k}\times \vec{k}_0|/k_0$ represent the magnitudes of longitudinal and transverse components of $\delta\vec{k}$ with respect to $\vec{k}_0$, respectively. Consequently, the wave packet $\bar{\phi}(\vec{x},t)$ can be approximated by
\begin{align}
  \bar{\phi}(\vec{x},t)&\simeq\frac{1}{(2\pi)^{3/2}}\exp(i\vec{k}_0\cdot\vec{x}-ik_0 t)\int^{+\infty}_{-\infty} \exp\left(-\frac{\delta k^2}{2\Delta k^2}+il\varphi\right)\nonumber\\
  &\times\exp(i\delta\vec{k}\cdot\vec{x}-i \delta k_\Vert t-i\frac{\delta k^2_\perp }{2k_0} t)d\delta k^3.
\end{align}

In the wave packet $\bar{\phi}(\vec{x},t)$, the Gaussian factor within the integral reveals rotational symmetry about $\vec{k}_0$. To exploit this symmetry, we adopt a cylindrical coordinate system with the $z$-axis aligned along $\vec{k}_0$. In this configuration, the volume element $d\delta k^3$ transforms into $\delta k_\perp d\delta k_\perp d\delta k_\Vert d\varphi$, and the term $\delta\vec{k}\cdot \vec{x}$ can be expressed as $\delta k_\Vert z + \delta k_\perp r_\perp \cos(\varphi-\varphi')$, where $r_\perp=\sqrt{x^2+y^2}$ represents the radial distance and $\varphi'$ the azimuthal angle of the position vector $\vec{x}$.  Consequently, the integration in the wave packet $\bar{\phi}(\vec{x},t)$ is reformulated as
\begin{align}\label{app-int}
  &\quad \int^{+\infty}_{-\infty} \exp\left(-\frac{\delta k^2_\Vert}{2\Delta k^2}\right)\exp(i\delta k_\Vert z-i \delta k_\Vert t) d\delta k_\Vert\nonumber\\
  & \times \int^{+\infty}_{0} \delta k_\perp \exp\left(-\frac{\delta k^2_\perp}{2\Delta k^2}-i \frac{\delta k^2_\perp t}{2k_0}\right)\int^{2\pi}_{0}\exp(i\delta k_\perp r_\perp\cos(\varphi-\varphi')+il\varphi) d\varphi d\delta k_\perp.
\end{align}

The equation \eqref{app-int} consists of two integrals. The first integral captures the longitudinal distribution of the wave packet through a standard Gaussian integral, yielding
\begin{equation}\label{app-int1}
  \int_{-\infty}^{+\infty} \exp\left(-\frac{\delta k^2_\Vert}{2\Delta k^2}\right)\exp(i\delta k_\Vert z-i \delta k_\Vert t) d\delta k_\Vert = \sqrt{2\pi}\Delta k\exp\left(-\frac{\Delta k^2(z-t)^2}{2}\right),
\end{equation}
which reflects the dispersion of the wave packet along its propagation axis.

In the second integral, the process begins with an inner integration over the azimuthal angle $\varphi$, solvable analytically to provide
\begin{equation}\label{app-int2}
  \int_{0}^{2\pi} \exp(i\delta k_\perp r_\perp\cos(\varphi-\varphi')-i l\varphi)d\varphi = 2\pi i^{|l|}\exp(il\varphi')J_{|l|}(\delta k_\perp r_\perp),
\end{equation}
where $J_n(x)$ denotes the Bessel function of the first kind. This step elucidates the wave packet's azimuthal structure, incorporating the effects of intrinsic OAM quantified by the quantum number $l$.

The subsequent outer integration over $\delta k_\perp$ can be simplifed to 
\begin{equation}
  \int_{0}^{+\infty} \delta k_\perp \exp\left(-\frac{\delta k^2_\perp}{2\Delta k^2}-i \frac{\delta k^2_\perp t}{2k_0}\right) J_{|l|}(\delta k_\perp r_\perp) d\delta k_\perp,
\end{equation}
resulting in a complex expression involving the modified Bessel functions of the first kind,
\begin{align}\label{app-int3}
  &\quad\int_{0}^{+\infty} \delta k_\perp \exp\left(-\frac{\delta k^2_\perp}{2\Delta k^2}-i \frac{\delta k^2_\perp t}{2k_0}\right) J_{|l|}(\delta k_\perp r_\perp) d\delta k_\perp\nonumber\\
  &= \frac{\sqrt{\pi} r_\perp\Delta k^3 k_0^{3/2}}{2\sqrt{2}(k_0+i\Delta k^2 t)^{3/2}}\exp\left(-\frac{r^2_\perp\Delta k^2 k_0}{4(k_0+i\Delta k^2 t)}\right)\nonumber\\
  &\quad\times\left(I_{\frac{|l|-1}{2}}\left(\frac{r^2_\perp\Delta k^2 k_0}{4(k_0+i\Delta k^2 t)}\right)-I_{\frac{|l|+1}{2}}\left(\frac{r^2_\perp\Delta k^2 k_0}{4(k_0+i\Delta k^2 t)}\right)\right),
\end{align}
where $I_a(x)$ signifies the modified Bessel function of the first kind. Incorporating the results from equations \eqref{app-int1}, \eqref{app-int2} and \eqref{app-int3} into equation \eqref{app-int}, the wave packet $\bar{\phi}(\vec{x},t)$ can be succinctly expressed as follows:
\begin{equation}
  \begin{split}
  \bar{\phi}(\vec{x},t)&\simeq 
  \frac{\sqrt{\pi} r_\perp\Delta k^4 k_0^\frac{3}{2}}{2\sqrt{2}(k_0+it\Delta k^2)^\frac{3}{2}}\exp\left(-\frac{r_\perp^2\Delta k^2 k_0^2}{4(k^2_0+t^2\Delta k^4)}-\frac{(z-t)^2\Delta k^2}{2}\right)\\
  &\times \left\{I_\frac{|l|-1}{2}\left(\frac{r_\perp^2\Delta k^2 k_0}{4(k_0+it\Delta k^2)}\right)-I_\frac{|l|+1}{2}\left(\frac{r_\perp^2\Delta k^2 k_0}{4(k_0+it\Delta k^2)}\right)\right\}\\
  &\times \exp\left(i\vec{k}_0\cdot \vec{x}-ik_0 t+il\varphi+it\frac{r_\perp^2\Delta k^4 k_0 }{4(k^2_0+t\Delta k^4)}\right).
  \end{split}
  \end{equation}

\section*{References}
\bibliographystyle{iopart-num}
\bibliography{reference.bib}

\end{document}